\documentclass{amsart}
\usepackage{amsaddr}
\usepackage{geometry}                
\usepackage{graphicx}

\begin{document}

\author{Katsuhiko Hamada}
\address[Katsuhiko Hamada]{The University of Tokyo, Tokyo, Japan}
\author{Hiromu Mori}
\address[Hiromu Mori]{Life Science Center of TARA, University of Tsukuba, Japan}
\author{Hiroyuki Shinoda}
\address[Hiroyuki Shinoda]{The University of Tokyo, Tokyo, Japan}
\author{Tomasz M. Rutkowski$^*$}\thanks{$^*$Corresponding author}
\address[Tomasz M. Rutkowski]{Life Science Center of TARA, University of Tsukuba, Japan \\
RIKEN Brain Science Institute, Japan}
\email[Corresponding author]{tomek@bci-lab.info}

\title{Airborne Ultrasonic Tactile Display Brain--computer Interface Paradigm} 

\maketitle

\markleft{K. HAMADA ET AL.}

\begin{abstract}
We study the extent to which contact--less and airborne ultrasonic tactile display (AUTD) stimuli delivered to the palms of a user can serve as a platform for a brain computer interface (BCI) paradigm. Six palm positions are used to evoke combined somatosensory brain responses, in order to define a novel contact-less tactile BCI. A comparison is made with classical attached vibrotactile transducers. Experiment results of subjects performing online experiments validate the novel BCI paradigm.
\end{abstract}

\tableofcontents

\section{Introduction}
\label{sect:introduction}

State--of--the--art brain computer interfaces (BCIs) are typically based on mental visual or auditory paradigms, as well as motor imagery paradigms, which require extensive user training and good eyesight or hearing. In recent years, alternative solutions have been proposed to make use of a tactile modality~\cite{sssrBCI2006,tactileBCIwaiste2010,HBCIscis2012hiromuANDtomek} to enhance brain--computer interfacing efficiency. The concept reported in this paper further extends the brain's somatosensory channel by the application of a contact--less stimulus generated with an airborne ultrasonic tactile display (AUTD)~\cite{autdFIRST2008}.
The rationale behind the use of the AUTD is that, due to its contact--less nature, it allows for a more hygienic application, avoiding the occurrence of skin ulcers (bedsores) in patients in a locked--in state (LIS).
This paper reports very encouraging results with AUTD--based BCI (autdBCI) in comparison with the classical paradigm of vibrotactile transducer--based somatosensory stimulus (vtBCI) attached to the user's palms ~\cite{HBCIscis2012hiromuANDtomek}.
The rest of the paper is organized as follows. The next section introduces the materials and methods used in the study. 
The results obtained in online experiments with $13$ healthy BCI users are then discussed. Finally, conclusions are formulated and directions for future research are outlined.

\section{Materials and Methods}

Thirteen male volunteer BCI users participated in the experiments. The users' mean age was $28.54$, with a standard deviation of $7.96$ years. The experiments were performed at the Life Science Center of TARA, University of Tsukuba, at the University of Tokyo and at RIKEN Brain Science Institute, Japan.
The online (real-time) EEG autdBCI and vtBCI paradigm experiments were conducted in accordance with the \emph{WMA Declaration of Helsinki - Ethical Principles for Medical Research Involving Human Subjects} and the procedures were approved and designed in agreement with the ethical committee guidelines of the Faculty of Engineering, Information and Systems at University of Tsukuba, Japan.
The AUTD stimulus generator produced vibrotactile contact--less stimulation of the human skin via the air using focused ultrasound ~\cite{autdFIRST2008,katsuhikoAUTD2014}.
The effect was achieved by generating an ultrasonic radiation static force produced by intense sound pressure amplitude (a
nonlinear acoustic phenomenon).
The radiation pressure deformed the surface of the skin on the palms, creating a tactile sensation.
An array of ultrasonic transducers mounted on the AUTD created the focussed radiation pressure at an arbitrary focal point by choosing a phase shift of each transducer appropriately (the so--called phased array technique).
Modulated radiation pressure created a sensation of tactile vibration similar to the one delivered by classical vibrotactile transducers attached to the user's palms.
The AUTD device developed by the authors~\cite{autdFIRST2008,katsuhikoAUTD2014} adhered to ultrasonic medical standards and did not exceed the permitted skin absorption levels (approximately $40$ times below the limits).
The effective vibrotactile sensation was set to $50$~Hz~\cite{katsuhikoAUTD2014}.
\begin{figure}[!t]
	\begin{centering}
	\includegraphics[width=\textwidth,clip]{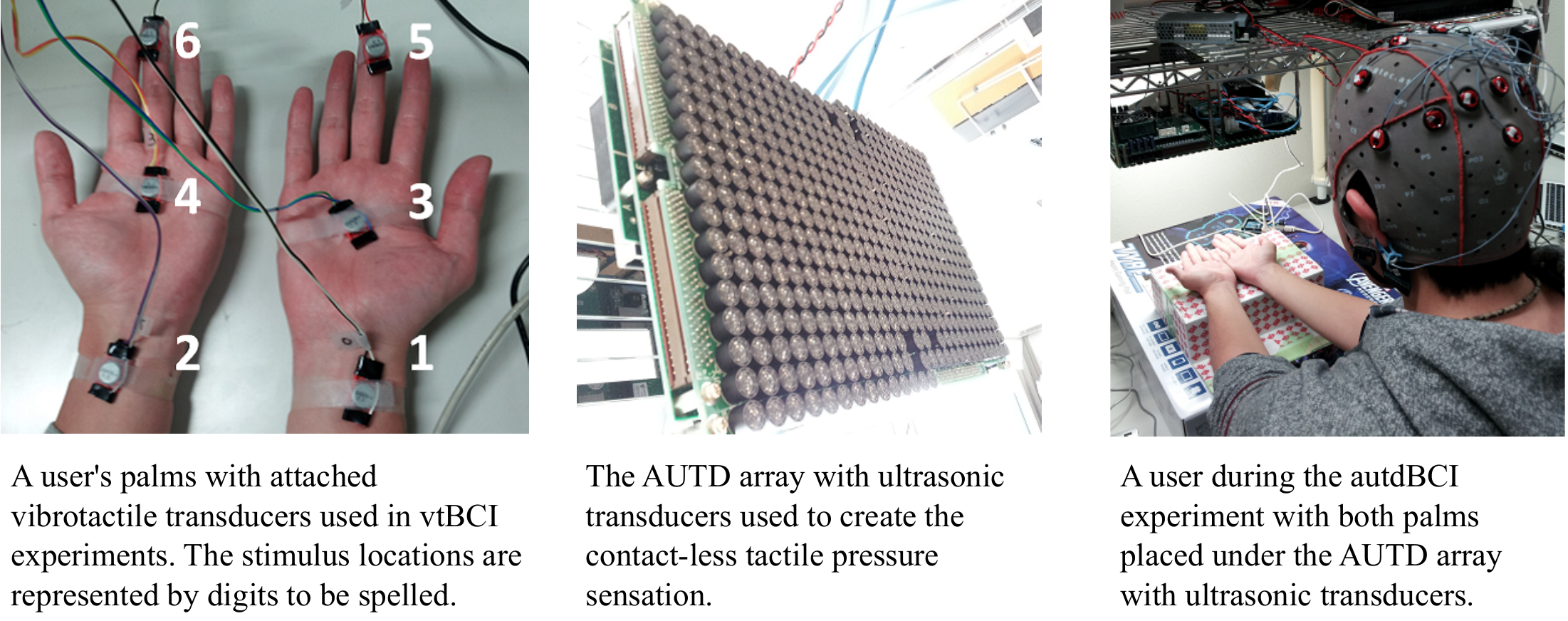}
	\caption{Tactile stimulus set-ups for autdBCI and vtBCI experiments.}
	\label{fig:expFIG}
	\end{centering}
\end{figure}
As a reference, in the second vtBCI experiment, contact vibrotactile stimuli were also applied to locations on the users' palms via the transducers HIHX09C005-8. Each transducer in the experiments was set to emit a square acoustic frequency wave at $50$~Hz, which was delivered from the ARDUINO micro--controller board with a custom battery--driven and isolated power amplifier and software developed in--house and managed from a MAX~6 visual programming environment.
The two experiment set-ups above are presented in Figure~\ref{fig:expFIG}.
Two types of experiments were performed with the volunteer users. Psychophysical experiments with button--press responses were conducted in order to test uniform stimulus difficulty levels from response accuracy and time measurements. The subsequent online BCI EEG experiments evaluated the autdBCI paradigm efficiency and allowed for a comparison with the classical skin contact--based vtBCI reference.
In both the above experiment protocols, the users were instructed to spell sequences of six digits representing the stimulated positions on their palms.
The training instructions were presented visually by means of the \emph{BCI2000} program~\cite{bci2000book} with the numbers $1-6$ representing the palm locations as depicted in the left panel of Figure~\ref{fig:expFIG}.
The EEG signals were captured with an EEG amplifier system g.USBamp by g.tec Medical Engineering GmbH, Austria, using $16$ active electrodes. The electrodes were attached to the head locations: \emph{Cz, Pz, P3, P4, C3, C4, CP5, CP6, P1, P2, POz, C1, C2, FC1, FC2,} and \emph{FCz}, as in the $10/10$ extended international system. The ground electrode was attached to the \emph{FPz} position, and the reference was attached to the left earlobe. No electromagnetic interference was observed from the AUTD or vibrotactile transducers operating with frequencies notch--filtered together with power line interference from the EEG.
The EEG signals captured were processed online with a BCI2000--based application~\cite{bci2000book}, using a stepwise linear discriminant analysis (SWLDA) classifier~\cite{krusienski2006} with features drawn from the $0-800$~ms ERP intervals.
The stimulus length and inter--stimulus--interval were set to $400$~ms, and the number of averages in BCI trials was set to $15$ in order to collect enough data for the classifier training. The EEG recording sampling rate was set at $512$~Hz, and the high and low pass filters were set at $0.1$~Hz and $60$~Hz, respectively. The notch filter to remove power line interference was set for a rejection band of $48 \sim 52$~Hz.
Each user performed three experiment sessions (randomized $90$~targets and $450$~non-targets each), which were later averaged for the online SWLDA classifier. 

\section{Results and Conclusions}\label{sec:results}

\begin{figure}[t]
	\begin{centering}
	\includegraphics[width=\textwidth,clip]{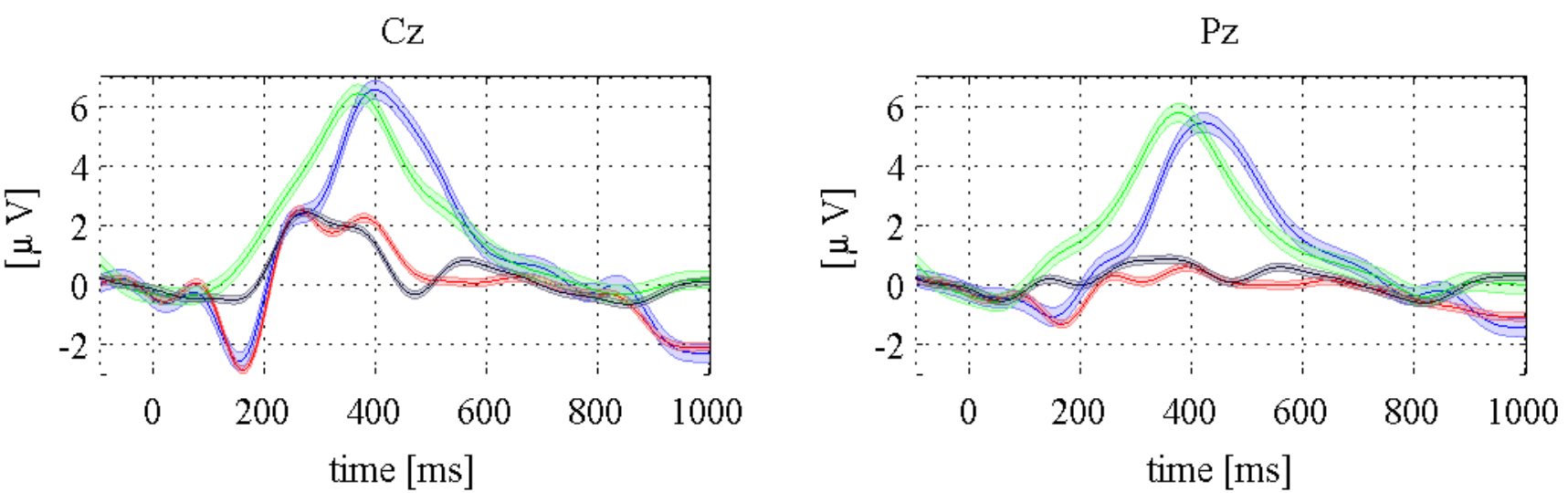}
	\caption{The autdBCI (blue - targets; red - non--targets) and vtBCI (green - targets; black - non--targets) grand mean averaged ERP responses, together with standard error bars. Due to limited space, of the $16$ measured, only electrodes \emph{Cz} and \emph{Pz} are presented.}
	\label{fig:erpFIG}
	\end{centering}
\end{figure}

\begin{figure}[b]
	\begin{centering}
	\includegraphics[width=\textwidth,clip]{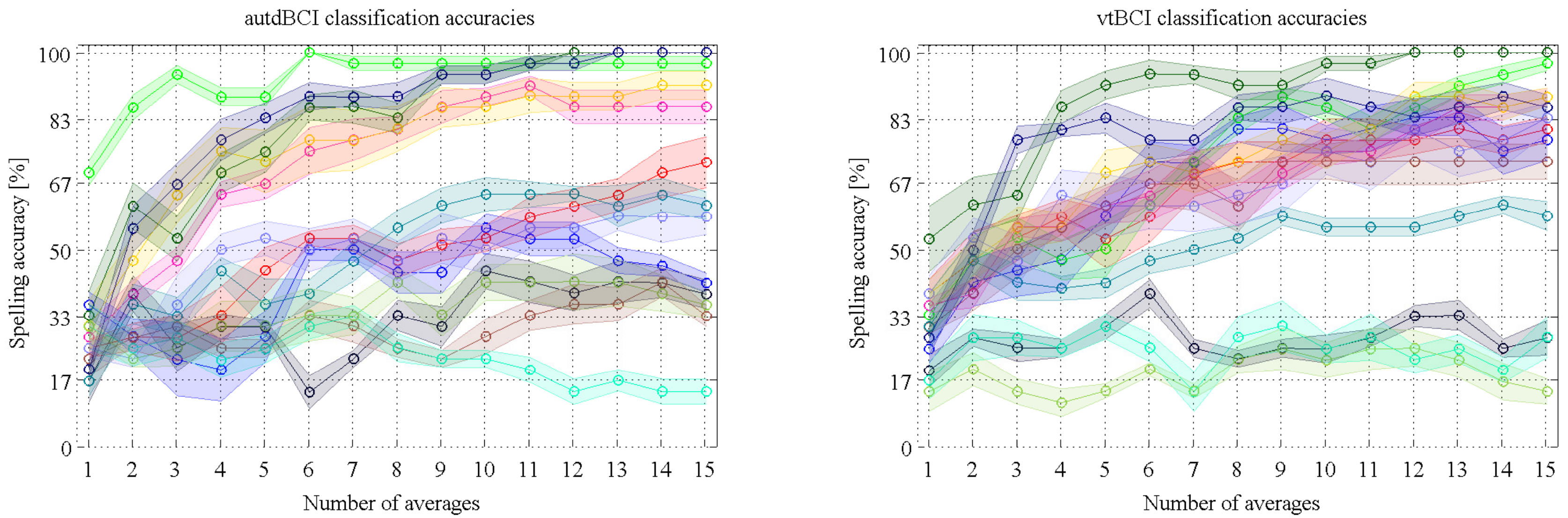}
	\caption{Averaged autdBCI and vtBCI six digits spelling accuracy results colour coded separately for each user, with standard error bars depicted.}\label{fig:accuracy}
	\end{centering}
\end{figure}

The averaged evoked responses to targets and non--targets are depicted together with standard error bars in Figure~\ref{fig:erpFIG}. The BCI six digit sequences spelling accuracy analyses for both the experiments for the various averaging options are summarized in Figure~\ref{fig:accuracy}. The chance level was of $16.6\%$. The mean accuracies for $15$-trial averaged ERPs were $63.8\%$ and $69.4\%$ for autdBCI and vtBCI, respectively. The maximum accuracies were $78.3\%$ and $84.6\%$ respectively. The differences were not significant, supporting the concept of AUTD--based tactile stimulus usability for BCI. However, a single trial classification offline analysis of the collected responses resulted in mean accuracies of $83.0\%$ for autdBCI and $53.8\%$ for vtBCI, leading to a possible $19.2$~bit/min and $7.9$~bit/min, respectively. 
In the case of the autdBCI, only a single user's results were bordering on the level of chance, and four subjects attained $100\%$ ($10$ trials averaging). On average, lower accuracies were obtained with the classical vtBCI, with which three users bordered on the level of chance, and only one user scored $100\%$ accuracy level in SWLDA--classified averaged responses.
This case study demonstrates results obtained with a novel six--command--based autdBCI paradigm. We compared the results with classical vibrotactile transducer stimuli already generated. The experiment results obtained in this study confirm the validity of the contact--less autdBCI for interactive applications and the possibility to further improve the results with the utilization of single trial--based linear classification.
The EEG experiment with the paradigm confirms that contact--less (airborne) tactile stimuli can be used to create six command--based interfaces.
The results presented offer a step forward in the development of novel neurotechnology applications. Due to the still not very high interfacing rate achieved by users in the case of online BCI, the current paradigm obviously requires improvement and modification. These requirements determine the major lines of study for future research. 

However, even in its current form, the proposed autdBCI can be regarded as a practical solution for LIS patients (locked into their own bodies despite often intact cognitive functioning), who cannot use vision or auditory-based interfaces due to sensory or other disabilities.

\section*{Author contributions} 
Designed and performed the EEG experiments: KH, HM, TMR. Analyzed the data: KH, HM, TMR. Conceived the concept of the AUTD--based BCI paradigm: TMR, HS. Supported the project: HS, TMR. Wrote the paper: KH, TMR.

\section*{Acknowledgements}

Hiromu Mori and Tomasz M. Rutkowski were supported in part by the Strategic Information and Communications R\&D Promotion Program (SCOPE) no. 121803027 of The Ministry of Internal Affairs and Communications in Japan.

%

\newcommand{\etalchar}[1]{$^{#1}$}

\end{document}